# Monitoring of nanoplasmonics-assisted deuterium production in a polymer seeded with resonant Au nanorods using in situ femtosecond laser induced breakdown spectroscopy

**N. Kroó[1,8], M. Aladi[1], M. Kedves[1], B. Ráczkevi[1], A. Kumari[1,4], P. Rácz[1], M. Veres[1], G. Galbács[1,2], L.P. Csernai[1,3,5], T.S. Biró[1,6,7]**

[1]HUN-REN Wigner Research Centre for Physics, NAPLIFE, Budapest, Hungary
[2]University of Szeged, Department of Molecular and Analytical Chemistry, Szeged, Hungary
[3]Csernai Consult Bergen, Ulset, Norway
[4]University of Marburg, Germany
[5]University of Bergen, Norway
[6]University Babes-Bolyai, Cluj, Romania
[7]Complexity Science Hub, Vienna, Austria
[8]Hungarian Academy of Sciences

*Abstract*

In this brief report, we present laser induced breakdown spectroscopy (LIBS) evidence of deuterium (D) production in a 3:1 urethane dimethacrylate (UDMA) and triethylene glycol dimethacrylate (TEGDMA) polymer doped with resonant gold nanorods, induced by intense, 40 fs laser pulses. The in situ recorded LIBS spectra revealed that the D/(2D+H) increased to 4-8% in the polymer samples in selected events. The extent of transmutation was found to linearly increase with the laser pulse energy (intensity) between 2 and 25 mJ (up to $3 \cdot 10^{17}$ W/cm²). The observed effect is attributed only to the field enhancing effects due to excited localized surface plasmons on the gold nanoparticles.

*Keywords:* fs-LIBS, nanoparticles, polymer, plasmonics, localized surface plasmons, gold

*Introduction*

Formerly we have observed laser-induced deuterium (D, or ²H) production in a 3:1 urethane dimethacrylate (UDMA) and triethylene glycol dimethacrylate (TEGDMA) polymer blend seeded with Au nanorods. The D production was indirectly detected by measuring the C-D oscillation mode by Raman spectroscopy on the laser ablation crater walls within the polymer [1]. The Au nanorods were resonant, for plasmonic excitation, with the used 795 nm Ti:Sa laser emitting a few tens of femtosecond (fs) long pulses with up to 25 mJ energy. The volume of the ablation craters has been also found to significantly increase in the nanorod-seeded polymer relative to the unseeded one using the same pulse energy/intensity, indicating at the highest laser intensities nuclear fusion energy production. This is attributed to plasmonic effects causing near field enhancement in the samples [2].

Since the 2D+H (where H denotes protium, or ¹H) content remains constant during a transmutation like $p + e + p \rightarrow d + \gamma + \nu$ (also known as the nuclear PEP process), the above finding is important and surprising. In order to obtain some more direct evidence about the increase of the D/(2D+2) ratio, hence D-production, atomic spectroscopy has been called in. In the present study, we employ laser-induced breakdown spectroscopy (LIBS). LIBS is a plasma atomic emission spectroscopy method typically performed using nanosecond laser pulses, and is widely used for the multi-element trace analysis of samples of various origin, also including samples related to fusion research [3-7]. The spectral lines recorded in the plasma emission can be used to determine the elemental composition of the sample. In our case, we used LIBS to detect the Balmer α lines of H and D, which are located around 656 nm, in the visible part of the spectrum. These electron transitions involve the *n*= 3 and the *n*= 2 states in the atoms, and their wavelength only differ by 0.18 nm. The LIBS method has a unique set of analytical features that can be taken advantage in the present project. It requires practically no sample preparation, provides single-shot detection and therefore consumes only pg to µg amount of sample, allows local analysis with micrometer-level spatial resolution. It can also be performed in a remote or stand-off manner, which is very useful in a vacuum setup [4, 6-9].

Moreover, it has been shown that the incorporation or deposition of metallic nanoparticles (NPs) in solid samples can greatly enhance the laser-matter interaction thereby producing higher signal intensities in LIBS [10, 11]. The interaction of femtosecond laser pulses with the sample has also been studied intensively in recent years in the LIBS literature [12-15].

In the present study, 40 fs laser pulses with different pulse energies ranging between 2.0 and 25 mJ were focused on UDMA-TEGDMA polymer targets placed inside a vacuum chamber. A microplasma is generated and the plasma light emission were collected by a high spectral resolution spectrograph in the vicinity of the H-α line at 656.29 nm and the D-α line at 656.11 nm. Thus, we use fs-LIBS to monitor the D production by the excitation laser pulse.

*Experimental details*

The LIBS experimental system used for the present study consists of a femtosecond Ti:Sa chirped-pulse amplifier laser with a central wavelength of 795 nm and maximum pulse energy of 30 mJ (Coherent Hidra). The laser pulse duration used for our experiment has been 40 fs. In our experiments singe shot operation was used. The peak intensity measured was $3 \cdot 10^{17}$ W·cm$^{-2}$. The laser induced breakdown plasma was generated by focusing the laser beam on the polymer target surface through one of the optical windows of the vacuum chamber, using a collection lens with a focal length of 40 cm and the focal plane was the same during the measurements. The plasma emission was collected at an angle of 45 degrees and collimated by placing a condenser lens at a 4 cm distance from the target inside the chamber. The collimated beam was then fed into a fiber collimator and then into a double Echelle high spatial resolution spectrometer equipped with an ICCD camera (Demon, LTB, Berlin) using a fiber-optic cable with a core diameter of 400 µm. The spectrometer can record spectra in the wavelength range of 190-900 nm with a spectral resolution of 2.5-12 pm. Light collection was done using a fixed gate width of 1 µs and the gate delay varied in the range of 0.1 µs to 1.0 µs. Based on the optimal signal and signal to noise level 0.6 µs delay was used during the evaluation of the presented in this paper measurements. The width of the inspection range (spectral window) of the spectrometer was 3 nm, now centered around 656 nm in order to observe the H-α and D-α lines. The schematic of the experimental arrangement is shown in Figure 1.

The laser not only triggered the spectrometer gating, but also a fast digital camera that was directed to observe the plasma plume from the side through another optical window on the vacuum chamber. in order to prevent overexposure, the camera was equipped with a narrow-band green blocking filter. The camera allowed us to observe the ablation plume and plasma. The polymer targets were fitted to a vacuum micropositioning stage operated manually from outside the chamber (manipulator), which allowed us to accurately position the targets with respect to the focused laser beam.

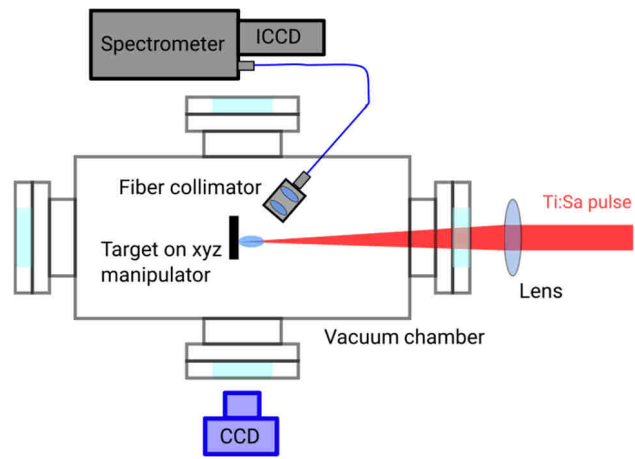

Fig. 1. A schematic of the experimental setup

Details of the preparation of the UDMA-TEGDMA polymer targets doped with resonant gold nanorods are described in [14]. Briefly, the procedure involved photopolymerization of the UDMA (Sigma Aldrich Co., USA) and TEGDMA (Sigma Aldrich Co., USA) monomers mixed and homogenized in a weight ratio of 3:1, after adding the photoinitiator camphorquinone (CQ, 0.2 m/m% concentration, obtained from Sigma Aldrich Co., USA), its co-initiator ethyl-4-dimethylaminobenzoate (EDAB, 0.4 m/m% concentration, obtained from Sigma Aldrich Co., USA), as well as the dodecanethiol-capped 25 x 75 nm gold nanorods (in a concentration of $1.9 \times 10^{12}$ mL$^{-1}$, which equals 0.108 m/m%) purchased from Nanopartz Inc., USA. The curing of the blend was done using a standard dental curing LED lamp (460 nm center wavelength).

For comparative purposes, another set of polymer targets were also photosynthesized using the same CQ-EDAB photoinitiator system, which contained UDMA and fully deuterated methyl methacrylate (d8-MMA) monomers. This resulted in polymers with covalently-bound, 23 m/m% D/(H+D) concentration.

All samples were analyzed in the vacuum chamber, filled up with argon gas to a 5 mbar pressure. Using a buffer gas is well established technique to enhance hydrogen LIBS signals [15, 16].

*Results*

In order to compensate for the differences in detection efficiency for H and D, due to the different propagation speed of these atoms within the ablation plume as well as to the slight difference in spectrometer sensitivities at the two detection wavelengths, we also carried out LIBS measurements on a UDMA-dMMA polymer target with known deuterium content. A measured typical LIBS spectrum is shown in Figure 2. The deuterium content in the sample was 23%, expressed as D/(D+H), while the ratio of the area of the spectral peaks ($A_{D-\alpha}/(A_{D-\alpha}+A_{H-\alpha})$) came out as 27.8 ± 2.4%. The ratio of these values were used as a correction factor when the D content produced by the fs laser pulses was calculated later (Figure 3 and 4).

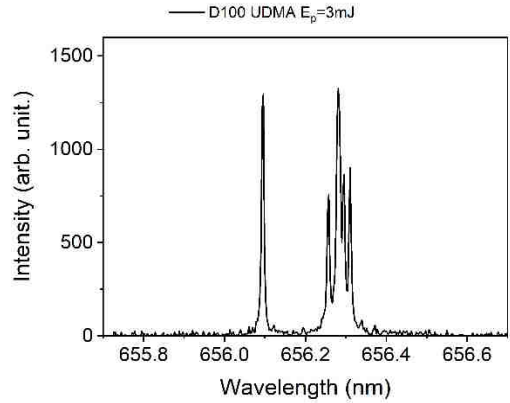

Fig. 2. Typical fs-LIBS spectrum of a partially deuterated UDMA-MMA sample.

Some typical single-shot LIBS spectra recorded during the incidence of fs laser pulses onto the UDMA-TEGDMA target are shown in Figure 3. Both the H-α line at 656.29 nm and the D-α line at 656.11 nm can be observed in the case of Au nanoparticle-seeded polymers (upper row), but for polymers without NPs the D line is mostly absent (lower row). This latter observation indicates that our LIBS arrangement is not sensitive enough to detect D at its natural D/H = 1/6002 ratio.

Please also note that due to the high amplification setting at the ICCD detector of the spectrometer needed for the detection of hydrogen lines, the baseline was noisy (shot noise) and some random, small intensity peaks also appeared in the vicinity – these peaks may also belong to trace metal contamination in the polymers originating from the high concentration organic components or the nanorods. Here we have to mention that, as was evidenced by light microscopy observations, our polymer target fabrication process could not, in spite of all our best efforts, produce completely scratch- and bubble-free polymers, therefore the laser beam could often hit these surface blemishes. This circumstance may further enhance the actual fraction of the observed D incidents.

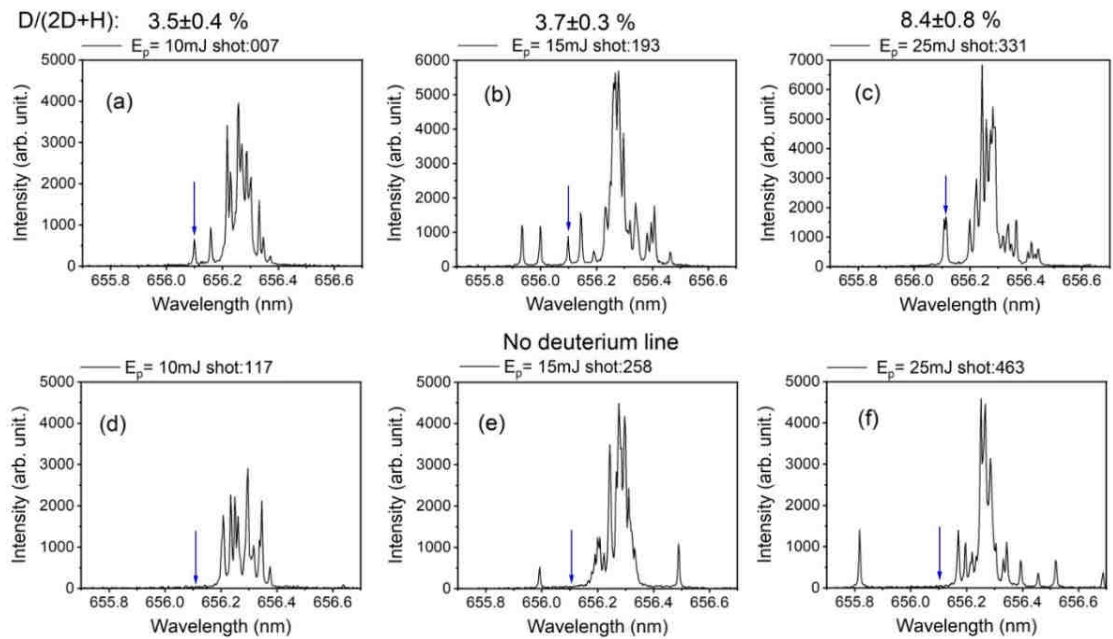

Fig. 3. fs-LIBS spectra with (a)-(c) and without (d)-(f) Au resonant nanorods, taken with 0.6 μs delay time after the laser shot. The intensity equivalents at the best focus with 20 μm diameter of the respective pulse energies are as follows: 25 mJ ≡ 3.0·10$^{17}$ W·cm$^{-2}$, 15 mJ ≡ 1.8·10$^{17}$ W·cm$^{-2}$, and 10 mJ ≡ 1.2·10$^{17}$ W·cm$^{-2}$. The arrows in the figures indicate the 656.11 nm position of the D-α spectral line. The laser pulse energy dependence of the deuterium production is also given.

The D generation was also found to enhance linearly with the increasing laser pulse intensity (Figure 3.), which also suggests that the effect can be induced by the EM field of the high intensity laser beam.

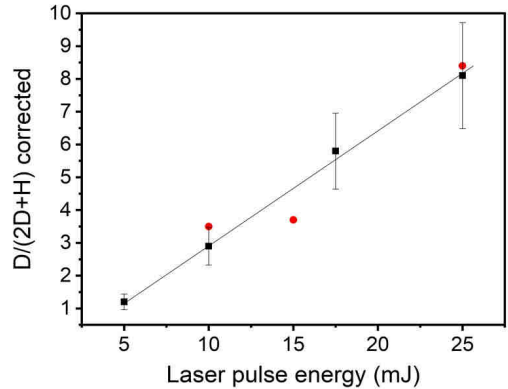

Fig. 4. Laser pulse energy dependence of the D/(2D+H) line intensities in percents. Two sets of the newly created deuterium atoms from two different measurement series are plotted. Data points are averages are based on calculations for 100 measurements. Standard deviations are represented as error bars, while red dots represent the cases a-c of Fig. 2

These results are surprising, but in accordance with the observations reported in [1] and [2]. Our numerical simulations also confirm the assumption that resonant nanorod antennas can accelerate protons [17], which lead to directed proton distribution [18]. This effect with adequate distribution of nanoantennas expected to achieve simultaneous ignition of nuclear processes [19] thus preventing Rayleigh-Taylor (RT) instabilities and pre-detonation.

**Conclusions**

The described in situ fs-LIBS data provide further confirmation of the deuterium production from hydrogen (protium) atoms by an intense fs laser pulse in a polymer, observed previously also by Raman spectroscopy [1]. The deuterium Balmer α line has been measured and correlated with the energy/intensity parameters of the exciting laser pulse. It has been found that the proportion of protium atoms transmuted to deuterium is unexpectedly high, it amounts to 4-8%. This can be attributed only to the field enhancing effects due to excited localized surface plasmons on the gold nanoparticles. The described results may be explained only by nuclear processes which have not yet been fully explored. We intend to further investigate the process in future by applying nuclear particle detecting techniques, too.

**Acknowledgements**

This work has been supported by the Hungarian Research Network and the National Research, Development and Innovation Office (NKFIH) of Hungary, via the projects: Nanoplasmonic Laser Inertial Fusion Research Laboratory NKFIH 2022-2.1.1-NL-2022-0002. L. P. Csernai acknowledges support from Wigner Research Center for Physics, Budapest. The authors would like to thank the Wigner GPU Laboratory at the Wigner Research Center for Physics for providing support in computational resources. This work was supported by several institutions and research projects, including the Frankfurt Institute for Advanced Studies (Germany), the former Roland Eötvös Research Network of Hungary, the Research Council of Norway (No. 255253). A special thanks goes out to NKFIH for the support provided within the frameworks of the following projects: Nanoplasmonic Laser Fusion Research Laboratory (NKFIH-874-2/2020 and NKFIH-468-3/2021), Optimized nanoplasmonics (K 116362), Ultrafast physical processes in atoms, molecules, nanostructures and biological systems (EFOP-3.6.2-16-2017-00005), Development of micro- and nanostructures for laser and plasma spectroscopy (K 146733) and the Thematic excellence program in materials science and photonics (TKP2021-NVA-19).